\documentclass[sn-mathphys]{sn-jnl}
\usepackage{amsmath,amssymb}
\jyear{2023}
\theoremstyle{thmstyleone}

\theoremstyle{thmstyletwo}

\theoremstyle{thmstylethree}

\usepackage[normalem]{ulem}
\usepackage{xcolor}

\graphicspath{{figs/}}

\begin{document}

\title[Brazilian Journal of Physics]{Mean-field criticality explained by random matrices theory}

\author*[1]{\fnm{Roberto} \sur{da Silva}}\email{rdasilva@if.ufrgs.br}

\author[1]{\fnm{Heitor} \sur{C. M. Fernandes}}\email{heitor@if.ufrgs.br}
\equalcont{These authors contributed equally to this work.}

\author[1]{\fnm{Eliseu} \sur{Venites Filho}}\email{eliseu.venites@ufrgs.com}
\equalcont{These authors contributed equally to this work.}

\author[1]{\fnm{Sandra} \sur{D. Prado}}\email{prado@if.ufrgs.br}
\equalcont{These authors contributed equally to this work.}

\author[2]{\fnm{J.R.} \sur{Drugowich de Felicio}}\email{drugo@usp.br}
\equalcont{These authors contributed equally to this work.}

\affil*[1]{\orgdiv{Instituto de F{\'i}sica}, \orgname{Universidade Federal do Rio Grande do Sul}, \orgaddress{\street{Av. Bento Gon{\c{c}}alves, 9500}, \city{Porto Alegre}, \postcode{91501-970}, \state{Rio Grande do Sul}, \country{Brazil}}}

\affil[2]{\orgdiv{Departamento de F\'{\i}sica}, \orgname{Faculdade de Filosofia Ci\^{e}ncias e Letras de
Ribeir\~{a}o Preto, Universidade de S\~{a}o Paulo}, \orgaddress{\street{Av. dos Bandeirantes 3900}, \city{Ribeirão Preto}, \postcode{10587}, \state{S\~{a}o Paulo}, \country{Brazil}}}


\abstract{How a system initially at infinite temperature responds when suddenly placed at finite temperatures 
is a way to check the existence of phase transitions. It has been shown in [R. da Silva, IJMPC 2023] that phase transitions are imprinted in the spectra of matrices built from time evolutions of magnetization of spin models. In this paper, we show that this method works very accurately in determining  the critical temperature in the mean-field Ising model. We show that for Glauber or Metropolis dynamics, the average eigenvalue has a minimum at the critical temperature, which is corroborated by an inflection at eigenvalue dispersion at this same point. Such transition is governed by a gap in the density of eigenvalues similar to short-range spin systems. 
We conclude that the thermodynamics of this mean-field system can be described by the fluctuations in the spectra of Wishart matrices which suggests a direct relationship between thermodynamic fluctuations and spectral fluctuations.    
}

\maketitle

\section{Introduction}

\label{sec:introduction}


Ising-like Hamiltonians \cite{Salinas} can be simply written as:

\begin{equation}
\mathcal{H}=-J\sum\limits_{\left\langle i,j\right\rangle
}s_{i}s_{j}-h\sum\limits_{i=1}^{N}s_{i} \, ,
\label{Eq:short_range_hamiltonian}
\end{equation}%
where $s_{i}=\pm 1$ (spin 1/2). Here $h$ is the external field that couples
with each spin, and $\left\langle i,j\right\rangle $ denotes the sum over
the nearest neighbors in a $d$-dimensional lattice. Each spin, placed in its
original lattice, is linked to other $z=2^{d}$ neighbors. The number of
links in the lattice is $\frac{Nz}{2}$ where factor 1/2 avoids double
counting.

A mean-field (MF) approximation considers that each spin $s_{i}$ interacts
with a kind of \emph{magnetic cloud} represented (composed?) by the average
magnetization of all other spins, that is: $\xi _{i}=$ $\frac{1}{N}%
\sum\limits_{j=1,j\neq i}^{N}s_{j}$. Thus, in this approximation, the
interacting term $H_{int}=-J\sum\limits_{\left\langle i,j\right\rangle
}s_{i}s_{j}$\ must be replaced by%
\begin{equation*}
\mathcal{H}_{int}^{(MF)}=-\frac{Jz}{2}\sum\limits_{i=1}^{N}s_{i}\xi
_{i}\approx -\frac{Jz}{2N}\sum\limits_{i=1}^{N}\sum%
\limits_{j=1}^{N}s_{i}s_{j}.
\end{equation*}

Finally, one has that mean-field Hamiltonian is given by 
\begin{equation}
\mathcal{H}^{(MF)}=-\frac{Jz}{2N}M^{2}-hM\text{,}
\label{Eq:long_hange_hamiltonian}
\end{equation}%
where $M=\sum\limits_{i=1}^{N}s_{i}$\ is the total magnetization of the
system. Its properties are well known and, once our interest is in its
behavior close to the continuous phase transition at $\beta_c z J =1$, the
external field $h$ is made null in this work.

To briefly contextualize, the long-range (LR) MF regime has been explored in
several models \cite{Wreszinski}. It has been also shown that Monte Carlo
(MC) simulations could be applied to the MF Ising Model \cite%
{Henriques,Drugowich-Libero}.

When dealing with equilibrium properties, we can choose any prescription for
system's dynamics that satisfies the detailed balance condition, for
example, the usual Glauber dynamics for a single spin flip, $s_i \to - s_i$: 
\begin{equation}
w(s\rightarrow s^{(j)})=\frac{1}{2\tau }\left[ 1-\tanh \left( \beta \frac{%
\Delta \mathcal{H}^{(MF)}}{2}\right) \right]
\end{equation}%
where $\tau $\ is a characteristic parameter that can be made equal to one
or fitted to the time scale of the specific problem, $\beta=1/(k_BT)$ is the
inverse of temperature ($T$) and $\Delta \mathcal{H}^{(MF)}$ is system's
energy change due to the flip of spin $s_i$.

In the case of the MF Ising model, one has $\Delta \mathcal{H}^{(MF)}=-\frac{%
J}{2N}z\left[ \left( \sum_{i=1,\ i\neq j}^{N}s_{i}-s_{j}\right) ^{2}-\left(
\sum_{i=1,\ i\neq j}^{N}s_{i}+s_{j}\right) ^{2}\right] =\frac{2J}{N}%
zs_{j}\left( M-s_{j}\right) \approx \frac{2J}{N}zs_{j}M$\ . Thus, since $%
s_{j}^{2}=1$, $\left\langle s_{j}\right\rangle =m$, and considering that in
the MF one has$\ \left\langle \tanh \left( \frac{\beta J}{N}zM\right)
\right\rangle =\tanh \left\langle \frac{\beta J}{N}zM\right\rangle =\tanh
(\beta zJm)$. With these quantities, it is possible to write the time
evolution of magnetization~\cite{MarioBook} 
\begin{equation}
\tau \frac{dm}{dt}=-m+\tanh {(\beta Jzm)}
\end{equation}%
where $m=\lim_{N\rightarrow \infty }\frac{\left\langle M\right\rangle }{N}$.
One should note that the RHS of this equation is exactly the negative of
free energy of the Ising model: 
\begin{equation*}
\tau \frac{dm}{dt}=\left. \frac{\partial f}{\partial y}\right\vert _{y=m},
\end{equation*}%
with $f(y)=\frac{\Phi (y,h=0)}{Jz}=\frac{y^{2}}{2}-\frac{1}{\beta Jz}\ln
(2\cosh (\beta Jzy))$. Surely, for $\beta_c Jz=1$ (critical point), $m<<1$,
thus $\tau \frac{dm}{dt}=-m+\tanh m\approx -\frac{1}{3}m^{3}$. And then at
criticality, the Ising model has the asymptotical decay given by \cite%
{MarioBook,Anteneodo,SilvaBJP2022}: 
\begin{equation}
m(t)=m_{0}\sqrt{\frac{3}{3+2m_{0}^{2}t/\tau }}\sim t^{-1/2} ~~~~~~~{%
\mbox{for }} t\rightarrow \infty  \label{Eq:Initial_magnetization_behavior}
\end{equation}%
where $m_0=m(t=0)$. Here it is important to mention that such behavior can
also be captured by performing time-dependent simulations in which $m(t)$ is
obtained for each $t-$th MC step, as an average over many different runs,
i.e., different time series of magnetization \cite{Anteneodo,SilvaBJP2022}.

On the other hand, short-range systems like the two-dimensional Ising model
at criticality has a behavior determined by short-time dynamics, a theory
that prescribes a crossover between two power-law: 
\begin{equation}
m(t)=\left\{ 
\begin{array}{lll}
m_{0}t^{\theta }, & ~~~\text{if}~~ & t<m_{0}^{z/x_{0}} \\ 
&  &  \\ 
t^{-\frac{\beta \nu }{z}}, & ~~~ \text{if} ~~ & m_{0}^{z/x_{0}}<t<t_{\infty }%
\text{.}%
\end{array}%
\right.  \label{Eq:shorttime}
\end{equation}
Here $t_{\infty }$\ is the equilibrium time, $\beta $\ and $\nu $\ are the
static exponents, while $z$\ is the dynamic one. The new exponent $\theta
=(x_{0}-\frac{\beta }{\nu })/z$\ governs the initial anomalous behavior of
magnetization, where $x_{0}$\ is known as the anomalous dimension of initial
magnetization. (Note that we use $\beta$ to denote the critical exponent to
maintain the tradition of the field. Unless explicitly specified, $%
\beta=1/k_BT$). The exponent $\theta $ can be obtained in two ways via
time-dependent MC simulations. First, by performing simulations with an
initial state prepared with fixed but random magnetization $m_{0}<<1$, and
thus one calculates $\theta $ considering an average to obtain $m(t)$ for
each $t-$MC step over tens of hundreds of different time evolutions and
extrapolating the result for $m_{0}\rightarrow 0$ \cite{Zheng}. In a second
way, by considering simulations with the initial random state also at $%
T\rightarrow \infty $ (spins chosen with probability 1/2). In this case, $%
m_{0}$ is not fixed, but by construction very small and with $\left\langle
m_{0}\right\rangle \approx 0$. The correlation is obtained by considering
the time correlation $C(t)=\frac{1}{N}\left\langle
\sum\limits_{i=1}^{N}\sum\limits_{j=1}^{N}s_{i}(t)s_{j}(0)\right\rangle $ ,
which, also, behaves as $C(t)\sim t^{\theta }$ such as $m(t)$ \cite%
{Tome,Tome2}.

For systems starting from $m_{0}=1$, one does not observe the initial slip
characterized by the exponent $\theta $ \cite{Zheng,Huse,Janssen}, instead
the magnetization decays with the second power law behavior $m(t)\sim t^{-%
\frac{\beta \nu }{z}}$\ directly, followed by an exponential decay at
thermodynamic equilibrium. Note that, for $T>T_{C}$\ or $T<T_{C}$, one does
not observe power laws at short times, and one must observe a stretched
exponential behavior for magnetization.

The exponent $\theta $ and consequently the initial slip of magnetization
for systems at high temperature ($m_{0}<<1$) is related to how the spin the
system reacts when suddenly placed at a finite temperature, more precisely
in this case at $T=T_{C}$.

Thus the short-time theory suscitates relevant questions about how the the
system captures the critical behavior or weak first-order transitions
behavior before thermalization \cite{rdasilva2014} even in nonequilibrium
models \cite{rdasilva2015,Silva2020,Hinchsen,Dickman,Pleimling}. These are
important questions since one can determine not just the critical exponents
but also localize the critical parameters (see, for example, a method that
we developed to optimize power-laws in \cite{SilvaPRE2012})

However, this signature of criticality out of equilibrium seems to be
inserted in ways even more notorious that can reflect what happens when
uncorrelated systems ($T\rightarrow \infty $) are placed at finite
temperatures, more importantly at $T\approx T_{C}$. Recently \cite{RMT2023},
by using random matrices built from time evolutions of magnetization in
earlier times of a spin system, we show how their spectra respond to phase
transitions. We showed how much the spectral properties of a statistical
mechanics system could be affected by criticality out of equilibrium. In
this case, we used the short-range two-dimensional Ising model as a test
model.

We also showed in this same work that by building such correlation random
matrices, known as Wishart matrices, from different time series of
magnetization simulated with MC, the density of eigenvalues of such matrices
can, with excellent precision, capture the phase transition of this system.
In another recent paper, we show that our method can go beyond responding
not only for critical points but also for strong first-order points \cite%
{RMT2023-2}

In this current contribution, we want to answer another question: can this
method be used for long-range systems such as the mean-field Ising system,
since such systems have different behavior in earlier times? The answer is
positive. To show that we build Wishart matrices for time evolutions of the
mean-field Ising model by considering MC simulations of such systems. We
will show that the random matrices method based on Wishart matrices can also
be used to describe the phase transition in this long-range system. We will
show that, similarly to the short-range systems, the method works very well
to localize the critical point of the MF Ising model.

In the next section, we present the necessary points of random matrices
theory necessary to contextualize the problem. In the sequence, we present
our main results in section \ref{Sec:Results} and finally in section \ref%
{Sec:Conclusions} a brief summary of our results and conclusions.

\section{Random matrices and spin systems}

We can assert that random matrices theory had its origin in the context of
nuclear physics, where E. Wigner \cite{Wigner,Wigner2} considered to
describe the complex energy levels of heavy-weight nucleus representing its
Hamiltonian by matrices with random entries.

If we consider symmetric ($h_{ij}=h_{ji}$) and well-behaved entries, i.e.,
distributed according to a probability density function $f(h)$ such that 
\begin{equation*}
\int_{-\infty }^{\infty }dh_{ij}f(h_{ij})h_{ij}^{k}<\infty \text{,}
\end{equation*}%
for $k=1,2$, of a matrix $H$, with dimension $N\times N$, and independent
entries, and therefore with joint distribution given by: 
\begin{equation*}
\Pr \left (\prod_{i<j}h_{ij} \right )=\prod_{i<j}f(h_{ij})
\end{equation*}%
will lead to jointly eigenvalues distribution $P(\lambda _{1},...,\lambda
_{N})$, such that its density of eigenvalues:

\begin{equation*}
\sigma (\lambda )=\int_{-\infty }^{\infty }...\int_{-\infty }^{\infty
}P(\lambda ,\lambda _{2},\lambda _{3},...,\lambda _{N})d\lambda
_{2}...d\lambda _{N}
\end{equation*}%
is universally described by semi-circle law \cite{Mehta,Soshnikov1998}:%
\begin{equation}
\sigma (\lambda )=\left\{ 
\begin{array}{l}
\frac{1}{\pi }\sqrt{2N-\lambda ^{2}}~~~\text{if\ }~~~\lambda ^{2}<2N \\ 
\\ 
0\ ~~~\text{if }~~~\lambda ^{2}\geq 2N%
\end{array}%
\right.   \label{Eq:wigner}
\end{equation}%
In the particular case that $f(h_{ij})=\frac{e^{-h_{ij}^{2}/2}}{\sqrt{2\pi }}
$, one has the Boltzmann weight: 
\begin{equation*}
P(\lambda _{1},...,\lambda _{N})=C_{N}\exp \left[ -\frac{1}{2}%
\sum\limits_{i=1}^{N}\lambda _{i}^{2}+\sum\limits_{i<j}\ln \left\vert
\lambda _{i}-\lambda _{j}\right\vert \right] 
\end{equation*}%
where $C_{N}^{-1}=\int_{0}^{\infty }...\int_{0}^{\infty }d\lambda
_{1}...d\lambda _{N}\exp [-\mathcal{H}(\lambda _{1}...\lambda _{N})]$,
corresponding to the Coulomb gas Hamiltonian:%
\begin{equation*}
\mathcal{H}(\lambda _{1}...\lambda _{N})=\frac{1}{2}\sum_{i=1}^{N}\lambda
_{i}^{2}-\sum_{i<j}\ln \left\vert \lambda _{i}-\lambda _{j}\right\vert 
\end{equation*}%
at temperature $\beta ^{-1}=1$. The last term is a logarithmic repulsion
exactly as the standard Wigner/Dyson \cite{Dyson} ensembles, while the first
is an attractive term. For both hermitian or symplectic entries \cite{Mehta}
the result is similar also resulting in $P(\lambda _{1},...,\lambda
_{N})=C_{N}^{(\beta )}\exp (-\beta \mathcal{H)}$ with respectively $\beta =2$
and $4$, and universally leading to the same density of eigenvalues from Eq.%
\ref{Eq:wigner}.

Despite this analogy, we do not have a direct connection between the
thermodynamics of a physical system and the fluctuations from random
matrices obtained from acquired data from this same physical system. This
connection emerges when we look at matrix correlations. With this knowledge,
we can recover the results from phase transitions and critical phenomena
from Thermostatistics. Surprisingly, only Wishart \cite{Wishart}, around
thirty years before Wigner and Dyson, focused on analyzing correlated time
series. Instead of using Gaussian or Unitary ensembles, he considered the
so-called Wishart ensemble, which essentially considers random correlation
matrices.

Thus, looking at such direction, we here define the main object for our
analysis, the magnetization matrix element $m_{ij}$ that denotes the
magnetization of the $j$-th time series at the $i$-th MC step of a system
with $N$ spins. Here $i=1,...,N_{MC}$, and $j=1,...,N_{sample}$. So the
magnetization matrix $M$ is $N_{MC}\times N_{sample}$. In order to analyze
spectral properties, an interesting alternative is to consider not $M$ but
the square matrix $N_{sample}\times $ $N_{sample}$:

\begin{equation*}
G=\frac{1}{N_{MC}}M^{T}M\ ,
\end{equation*}%
such that $G_{ij}=\frac{1}{N_{MC}}\sum_{k=1}^{N_{MC}}m_{ki}m_{kj}$, known as
Wishart matrix \cite{Wishart}. At this point, instead of working with $%
m_{ij} $, it is more convenient to take the Matrix $M^{\ast }$, defining its
elements by the standard variables:%
\begin{equation*}
m_{ij}^{\ast }=\frac{m_{ij}-\left\langle m_{j}\right\rangle }{\sqrt{%
\left\langle m_{j}^{2}\right\rangle -\left\langle m_{j}\right\rangle ^{2}}},
\end{equation*}%
where: 
\begin{equation*}
\left\langle m_{j}^{k}\right\rangle =\frac{1}{N_{MC}}%
\sum_{i=1}^{N_{MC}}m_{ij}^{k}\ .
\end{equation*}

Thereby: 
\begin{equation}
\begin{array}{lll}
\displaystyle G_{ij}^{\ast } & = & \displaystyle\frac{1}{N_{MC}}\displaystyle%
\sum_{k=1}^{N_{MC}}\frac{m_{ki}-\left\langle m_{i}\right\rangle }{\sqrt{%
\left\langle m_{i}^{2}\right\rangle -\left\langle m_{i}\right\rangle ^{2}}}%
\frac{m_{kj}-\left\langle m_{j}\right\rangle }{\sqrt{\left\langle
m_{j}^{2}\right\rangle -\left\langle m_{j}\right\rangle ^{2}}} \\ 
&  &  \\ 
& = & \displaystyle\frac{\left\langle m_{i}m_{j}\right\rangle -\left\langle
m_{i}\right\rangle \left\langle m_{j}\right\rangle }{\sigma _{i}\sigma _{j}}%
\end{array}
\label{Eq:Correlation}
\end{equation}%
where $\left\langle m_{i}m_{j}\right\rangle =\frac{1}{N_{MC}}%
\sum_{k=1}^{N_{MC}}m_{ki}m_{kj}$ and $\sigma _{i}=\sqrt{\left\langle
m_{i}^{2}\right\rangle -\left\langle m_{i}\right\rangle ^{2}}$.
Analytically, if $m_{ij}^{\ast }$ are uncorrelated random variables, the
jointly distribution of eigenvalues is described by the Boltzmann weight 
\cite{GuhrPR,Seligman3}: 
\begin{equation*}
\begin{array}{lll}
P(\lambda _{1},...,\lambda _{N_{sample}}) & = & C_{N_{sample}}\exp \left[ -%
\frac{N_{MC}^{2}}{2N_{sample}}\sum_{i=1}^{N_{sample}}\lambda _{i}\right.
\bigskip \\ 
&  &  \\ 
&  & \left. -\frac{1}{2}\sum_{i=1}^{N_{sample}}\ln \lambda
_{i}+\sum_{i<j}\ln \left\vert \lambda _{i}-\lambda _{j}\right\vert \right]%
\end{array}%
\end{equation*}%
where $C_{N_{sample}}^{-1}=\int_{0}^{\infty }...\int_{0}^{\infty }d\lambda
_{1}...d\lambda _{N_{sample}}\exp [-\mathcal{H}(\lambda _{1}...\lambda
_{N_{sample}})]$, corresponding to the Hamiltonian:%
\begin{equation*}
\mathcal{H}(\lambda _{1}...\lambda _{N_{sample}})=\frac{N_{MC}^{2}}{%
2N_{sample}}\sum_{i=1}^{N_{sample}}\lambda _{i}+\frac{1}{2}\sum_{i=1}^{N_{sample}}\ln
\lambda _{i}-\sum_{i<j}\ln \left\vert \lambda _{i}-\lambda _{j}\right\vert
\end{equation*}%
the density of eigenvalues $\sigma (\lambda )$ of the matrix $G^{\ast }=%
\frac{1}{N_{MC}}M^{\ast T}M^{\ast }$ follows in this case the known
Marcenko-Pastur distribution \cite{Marcenko}, which is written as:

\begin{equation}
\sigma (\lambda )=\left\{ 
\begin{array}{l}
\dfrac{N_{MC}}{2\pi N_{sample}}\dfrac{\sqrt{(\lambda -\lambda _{-})(\lambda
_{+}-\lambda )}}{\lambda }\ ~~~ \text{if\ } ~~~\lambda _{-}\leq \lambda \leq
\lambda _{+} \\ 
\\ 
0\ \ \ \ \text{otherwise}%
\end{array}%
\right.  \label{Eq:MP}
\end{equation}%
where $\lambda _{\pm }=1+\frac{N_{sample}}{N_{MC}}\pm 2\sqrt{\frac{N_{sample}%
}{N_{MC}}}.$

Now our aim is to analyse the behavior of $\sigma _{\text{numerical}%
}(\lambda )$ considering $m_{ij}$ obtained from mean-field Ising model
simulated at different temperatures. We expect that when $T\rightarrow
\infty $, $\sigma _{\text{numerical}}(\lambda )$ must be closer to $\sigma
(\lambda )$ according to Eq. \ref{Eq:MP}. However, the interesting results
are for $T\approx T_{C}$ or $T<T_{C}$. They will be presented in the next
section.

\section{Results}

\label{Sec:Results}

We performed MC simulations of the mean-field Ising model for $N=10^{4}$
spins. Thus we obtained matrix elements $m_{ij}$, for $i=1,...,N_{MC}=300$
MC steps, and $j=1,...,N_{sample}=100$ different time series for each
temperature. We start with random configurations $m_{0}\approx 0$ ($%
T\rightarrow \infty $), and used Glauber dynamics except when explicitly
mentioned.

First, we show 20 different time series for sake of comparison. Only for
comparison, we also simulated a two-dimensional Ising model keeping the same
number of spins $L^{2}=10^{4}$. The evolutions are shown in Fig. \ref%
{Fig:Time_evolution}. Both systems were simulated in their respective
critical temperatures: $\frac{k_{B}T}{J}=\frac{2}{\ln (1+\sqrt{2})}$ and $%
\frac{k_{B}T}{Jz}=1$. 

\begin{figure}[tbp]
\begin{center}
\includegraphics[width=0.8\columnwidth]{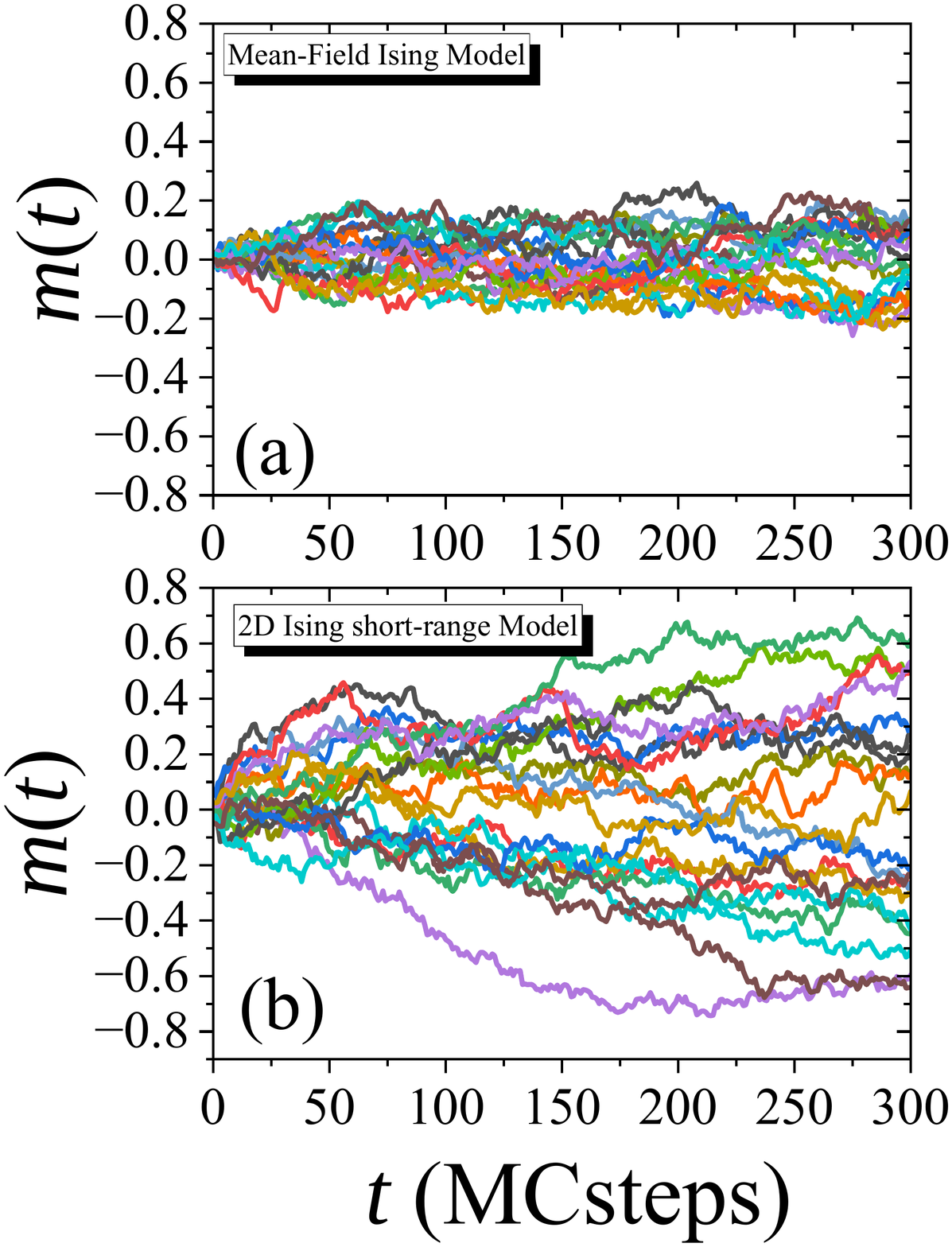}
\end{center}
\caption{A sample of magnetization time evolutions to be used for obtaining
the matrices and thus, performing the spectral study. Plot (a) shows the
series of mean-field Ising model that is the subject of this current study.
Plot (b) shows (for comparison) the results for the two-dimensional Ising
model. Both systems are simulated in their respective critical temperatures}
\label{Fig:Time_evolution}
\end{figure}

A sample of magnetization time evolutions to be used for obtaining the
matrices and thus, performing the spectral study. Plot (a) shows the series
of mean-field Ising model that is the subject of this current study. Plot
(b) shows (for comparison) the results of the two-dimensional Ising model.
We observe that the mean-field series seems to be present less variability
than the time series of the two-dimensional short-range Ising model. But the
question persists: should matrices $G$ built from MF Ising data produce
spectra that can describe the thermodynamics of this model?

Thus we build our ensemble of matrices, considering $N_{run}=1000$ different
matrices $G$ ($N_{sample}\times N_{sample}$). Then, we diagonalize them and
categorize the data between $\lambda _{\min }^{(\text{Numerical})}$ and $%
\lambda _{\max }^{(\text{Numerical})}$ among all eigenvalues, keeping the
number of bins fixed in $N_{bin}=100$.

We repeated the process for several different temperatures between $T_{\min
}=\frac{1}{2}T_{C}$ until $T_{\max }=\frac{13}{2}T_{C}$, recalling that $%
T_{C}=Jz/k_{B}$ (which is exactly 1 in our reduced units). We show the
numerical density of eigenvalues for the different temperatures in (Fig. \ref%
{Fig:density_of_eigenvalues}). 
\begin{figure}[tbp]
\begin{center}
\includegraphics[width=1.0\columnwidth]{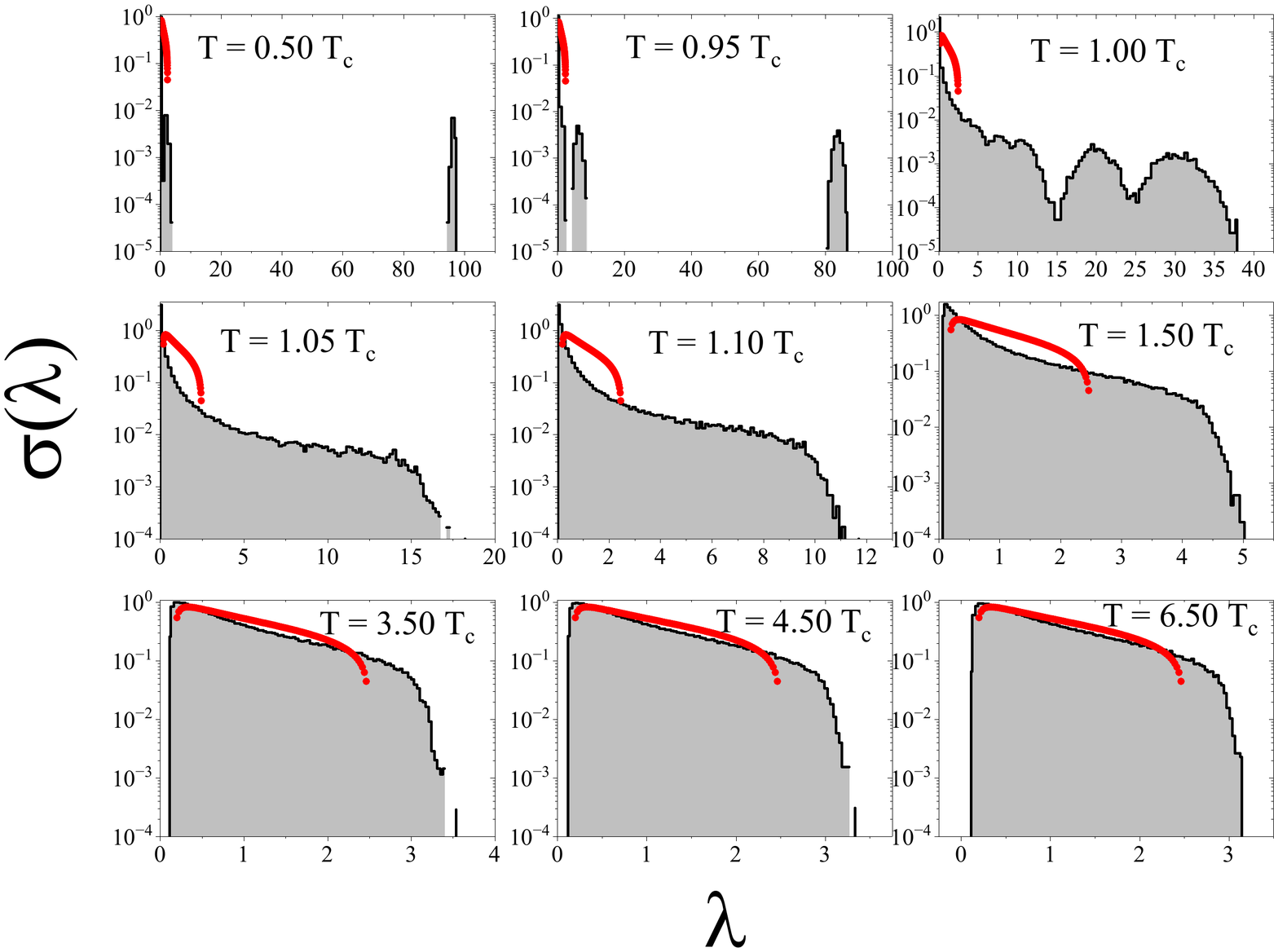}
\end{center}
\caption{Density of eigenvalues for different temperatures. We observe that
for $T=T_{C}$, there is no gap between two groups of eigenvalues that occurs
for $T<T_{C}$. For $T>T_{C}$ the density approaches to Marchenko-Pastur law
(Eq. \protect\ref{Eq:MP}) indicates that magnetization time series become
uncorrelated. The red continuous curve corresponds to Marchenko-Pastur
density for comparison.}
\label{Fig:density_of_eigenvalues}
\end{figure}
We observe that for exactly $T=T_{C}$, there is no gap between two groups of
eigenvalues that occurs for $T<T_{C}$. For $T>T_{C}$, the density approaches
to Marchenko-Pastur law (Eq. \ref{Eq:MP}), indicating that magnetization
time-series become uncorrelated.

Now, the first step of our study is ready. The spectra responded to the the
temperature of the system and apparently the gap between the eigenvalues
reduces to a single bulk. This also occurs with two-dimensional Ising and
Potts model (see \cite{RMT2023,RMT2023-2}). It is worth emphasizing that in
a direct comparison between these short-range systems with the current MF
one, the gap closing occurs exactly at $T=T_{C}$ while for the former, we
need to have a temperature a little higher than $T_{C}$ ($T\approx 1.10T_{C}$%
).

\begin{figure}[tbp]
\begin{center}
\includegraphics[width=1.0\columnwidth]{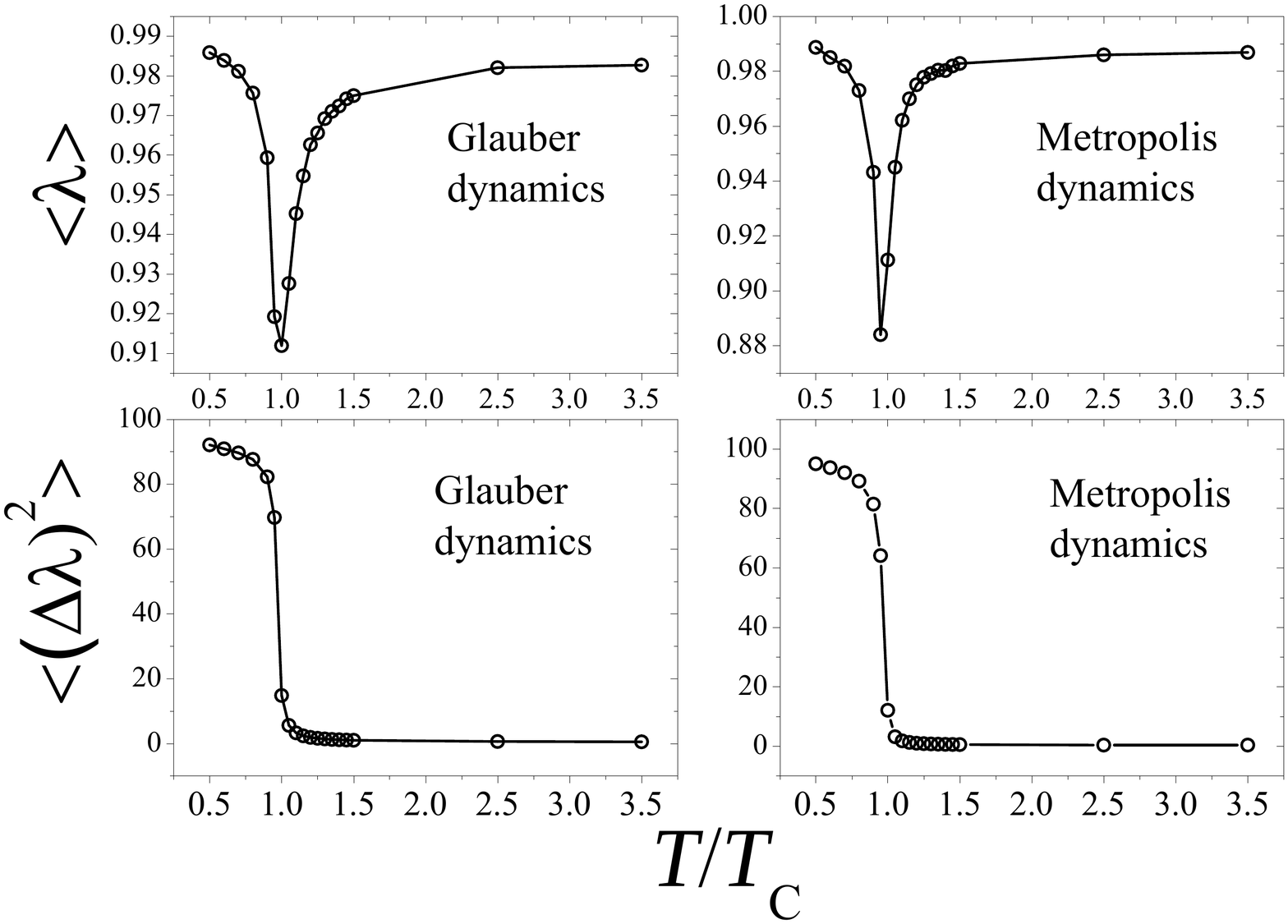}
\end{center}
\caption{Average eigenvalue and variance of the eigenvalue as a function of $%
T/T_{C}$. We observe a minimum on the average eigenvalue at $T=T_{C}$,
concurrently with an inflection of the variance at the same temperature. }
\label{Fig:fluctuations}
\end{figure}

However, independently of that, when we look at eigenvalues fluctuations,
our method is precise in asserting where is the critical temperature as
shown in Fig. \ref{Fig:fluctuations}. The results were obtained via MC
simulations from Glauber and Metropolis dynamics and they show very good
agreement.

By estimating the numerical moments:

\begin{equation*}
\left \langle\lambda ^{k}\right\rangle =\frac{\sum\limits_{i=1}^{N_{bins}}%
\lambda _{i}^{k}\sigma ^{(\text{Numerical)}}(\lambda _{i})}{%
\sum\limits_{i=1}^{N_{bins}}\sigma ^{(\text{Numerical)}}(\lambda _{i})}\text{%
,}
\end{equation*}%
we observe in the Fig. \ref{Fig:fluctuations}, the average eigenvalue as as
function of $T/T_{C}$, as well as the dispersion $\left\langle (\Delta
\lambda )^{2}\right\rangle =\left\langle \lambda ^{2}\right\rangle
-\left\langle \lambda \right\rangle ^{2}$ as function of the same quantity.
We performed MC simulations in this case for both Glauber and Metropolis
dynamics. We observe a minimal of the average eigenvalue at $T=T_{C}$,
concurrently with an inflection of the variance at the same point.

\begin{figure}[t]
\begin{center}
\includegraphics[width=0.8\columnwidth]{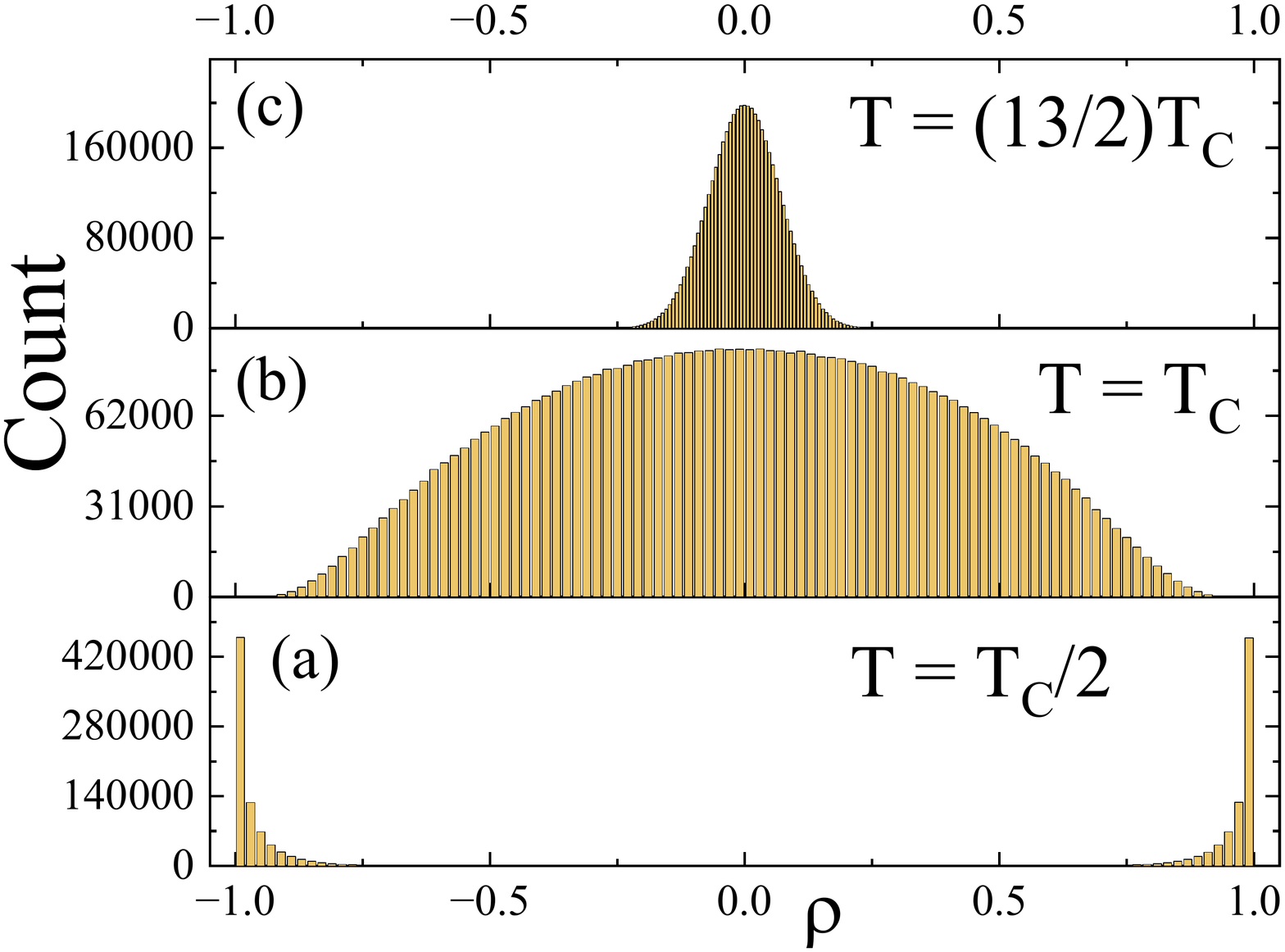}
\end{center}
\caption{Histograms of the correlations between different time evolutions
for the Mean-field Ising. Such study was performed for three different
temperatures, starting from random initial configurations. (a) $T=T_{C}/2$,
(b) $T=T_{C}$, and (c) $T=6T_{C}/5$, showing strong correlation,
indetermination, and decorrelation respectively, exactly as occurs in
short-range systems studied with the same method.}
\label{Fig:Correlation}
\end{figure}

Thus we can conclude that for this MF regime, the spectra of eigenvalues of
Wishart matrices built from magnetization time series work exactly as the
short-range models. However, it is interesting to investigate what occurs
directly on the correlations between the time series. Thus we build
histograms of elements of matrices $G$ directly in different temperatures
which can observe in Fig \ref{Fig:Correlation}.

Magnetization has a growth trend over the different evolutions when the
system suddenly quenches to a temperature $T=\frac{1}{2}T_{C}$. This the
tendency of ordering generates correlations between the different evolutions
corroborated by Fig. \ref{Fig:Correlation} (a), where considerable negative
or positive correlations occur depending on the initial configuration.

However, when the system quenches to a temperature $T=T_{C}$\ (Fig. \ref%
{Fig:Correlation} (b)), we see that correlation distribution is broadly
distributed. This is an intermediate situation that characterizes the
spontaneous breaking of symmetry.

The system becomes uncorrelated (no novelty) when the relaxation occurs from
a high temperature $T=\frac{13}{2}T_{C}$\ (Fig. \ref{Fig:Correlation} (c)).
Such qualitative behavior reflects on the spectra of Wishart matrices, whose
fluctuations can determine with precision the critical point of the system

\section{Conclusions and summaries}

\label{Sec:Conclusions}

Our purpose here was to apply a method of studying the spectra of
eigenvalues of Wishart matrices of magnetization time series for a
long-range spin system: the mean-field Ising model. Exactly as applied in
previous contributions (see refs. \cite{RMT2023,RMT2023-2}) in short-range
systems, we showed here that the method can also be applied to a mean-field
system with time series obtained from Monte Carlo simulations.

We believe that the method is promising and that it should be tested in
strict long-range systems, other magnetic spin systems, and nonequilibrium
models, among others.

It will be interesting to study the use of Wishart matrices in unsupervised
learning as those proposed, for example, in Refs.~\cite{wang2016,tirelli2022}%
.

\section*{Compliance with Ethical Standards}

\textit{Funding}: R. da Silva thanks CNPq for financial support under the
grant numbers 311236/2018-9 and 304575/2022-4.\newline

\textit{Conflict of Interest}: The authors declare that they have no known
competing financial interests or personal relationships that could have
appeared to influence the work reported in this paper.\newline

\textit{Ethical Conduct}: The authors declare that they did not violate any
ethical conduct in preparing this paper.

\end{document}